\documentclass[aps,pra,twocolumn,superscriptaddress,floatfix,showpacs,citeautoscript,longbibliography,hyperlinks]{revtex4-2}
\usepackage{amsmath}
\usepackage{epstopdf}
\usepackage{amsmath}
\usepackage{amssymb}
\usepackage{amsfonts}
\usepackage[colorlinks=true,citecolor=blue]{hyperref}
\usepackage{graphicx,graphics}

\usepackage{float}
\usepackage[caption = false]{subfig}

\newcommand{\cum}[2]{\langle \!\langle #1^{#2} \rangle\! \rangle}

\begin{document}
\title{Lee-Yang theory of Bose-Einstein condensation}
\author{Fredrik Brange}
\affiliation{Department of Applied Physics, Aalto University, 00076 Aalto, Finland}
\author{Tuomas Pyhäranta}
\affiliation{Department of Applied Physics, Aalto University, 00076 Aalto, Finland}
\author{Eppu Heinonen}
\affiliation{Department of Applied Physics, Aalto University, 00076 Aalto, Finland}
\author{Kay Brandner}
\affiliation{School of Physics and Astronomy, University of Nottingham, Nottingham NG7 2RD, United Kingdom}
\author{Christian Flindt}
\affiliation{Department of Applied Physics, Aalto University, 00076 Aalto, Finland}

\begin{abstract}
Bose-Einstein condensation happens as a gas of bosons is cooled below its transition temperature, and the ground state becomes macroscopically occupied. The phase transition occurs in the thermodynamic limit of many particles. However, recent experimental progress has made it possible to assemble quantum many-body systems from the bottom up, for example, by adding single atoms to an optical lattice one at a time. Here, we show how one can predict the condensation temperature of a Bose gas from the energy fluctuations of a small number of bosons. To this end, we make use of recent advances in Lee-Yang theories of phase transitions, which allow us to determine the zeros and the poles of the partition function in the complex plane of the inverse temperature from the high cumulants of the energy fluctuations. By increasing the number of bosons in the trapping potential, we can predict the convergence point of the partition function zeros in the thermodynamic limit, where they reach the inverse critical temperature on the real axis. Using less than 100 bosons, we can estimate the condensation temperature for a Bose gas in a harmonic potential in two and three dimensions, and we also find that there is no phase transition in one dimension as one would expect.
\end{abstract}

\maketitle

\section{Introduction}

Bose-Einstein condensation is a remarkable physical phenomenon by which a gas of bosons abruptly condenses into its ground state as it is cooled below its transition temperature \cite{Cornell:2002,pethick_smith_2008}. Bose-Einstein condensation was first observed with dilute atomic vapors that were cooled below their transition temperature of a few hundred nanokelvins \cite{Anderson:1995,Davis:1995}. These results spurred a wide range of developments that seek to explore how  Bose-Einstein condensates can be exploited for technological applications,~e.~g.~for quantum computing \cite{PhysRevA.85.040306}, sensing \cite{Aguilera_2014}, or thermal machines \cite{Myers2022,Eglinton2022}. At its core, Bose-Einstein condensation can be understood using basic arguments which relate the critical temperature of a gas solely to the density of bosons, $\rho$, the thermal de Broglie wavelength, $\lambda_{\mathrm{th}}=h/\sqrt{2\pi mk_B T}$, and the geometry and dimension of the confining potential. For example, for free bosons in three dimensions, the transition temperature can be found from the relation, $\rho \lambda_{\mathrm{th}}^3=\zeta(3/2)\simeq 2.61$, where $\zeta(x)$ is Riemann's zeta function \cite{Annett2004}. Typically, to observe Bose-Einstein condensation, the gas must be dilute, and the temperature must be ultralow. However, in recent years, Bose-Einstein condensation has also been observed at room temperature with quasiparticles, such as surface plasmons \cite{Deng2010,Hakala2018} or magnons \cite{Giamarchi2008,Demokritov2006}. To this end, a large de Broglie wavelength can be obtained by using bosons with a very low effective mass, rather than cooling them to subkelvin temperatures.

\begin{figure}[b]
    \centering
    \includegraphics[width=0.97\columnwidth]{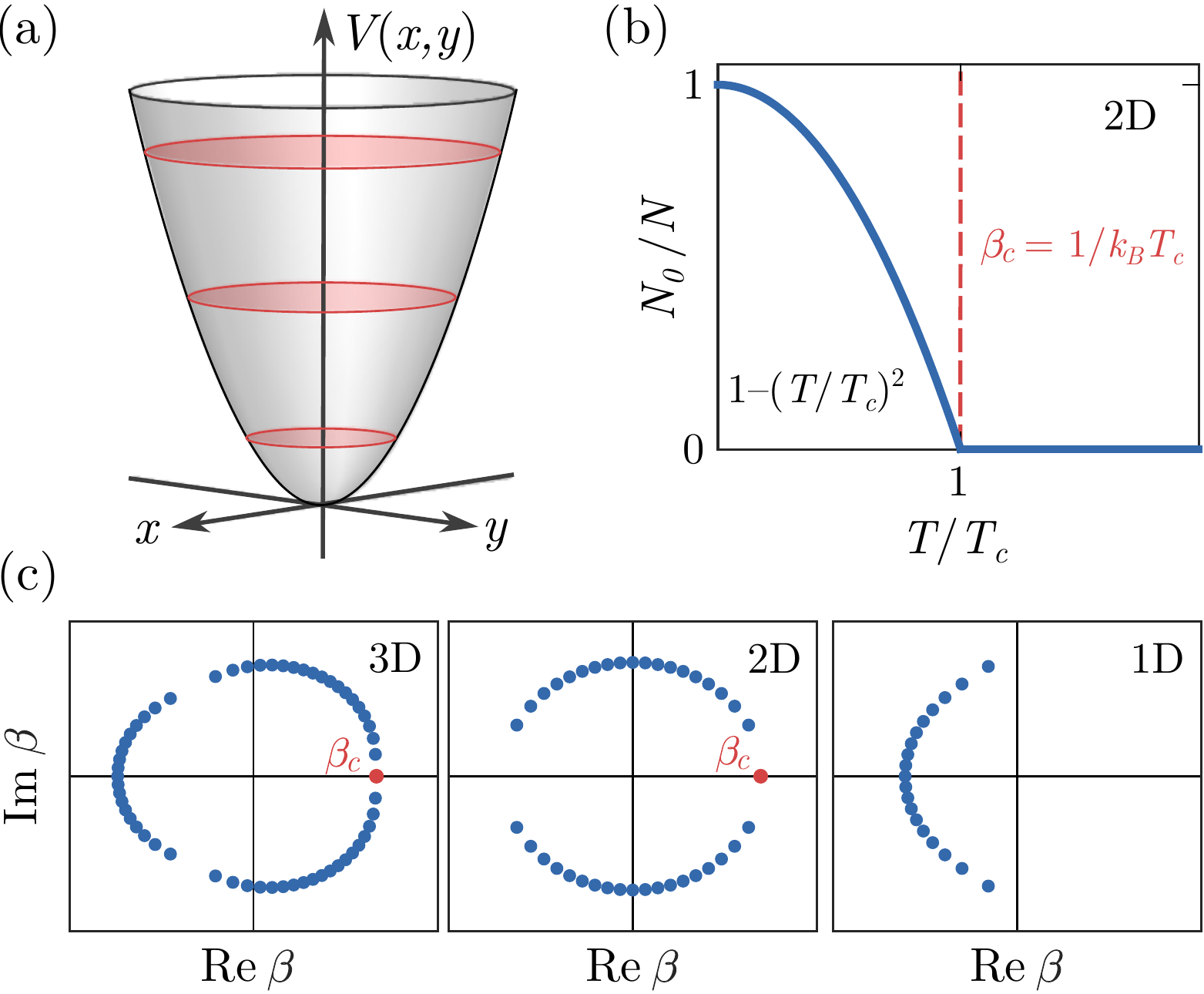}
    \captionsetup{justification=justified,singlelinecheck=false}
    \caption{Lee-Yang theory of Bose-Einstein condensation. (a) A two-dimensional harmonic potential $V(x,y)$ containing a gas of bosons. (b) Fraction of bosons in the ground state of a two-dimensional potential as a function of the temperature $T$. Below the critical temperature, $T_c$, the ground state becomes macroscopically occupied. (c) Zeros of the canonical partition function in the complex plane of the inverse temperature, $\beta = 1/(k_BT)$, for different dimensions. The zeros correspond to $N = 20$ bosons and were obtained with a high-temperature expansion of the partition function. The red dots illustrate potential convergence points $\beta_c$ in the thermodynamic limit.}
    \label{Fig1}
\end{figure}

Bose-Einstein condensation is a prime example of a phase transition, which can be analyzed within the framework of equilibrium statistical physics~\cite{Chandler1987,Goldenfeld1992,Kardar2007}. Phase transitions occur in the thermodynamic limit of many particles and large system sizes, and they are signaled by a nonanalytic behavior of the free-energy density at the critical value of the external control parameter, for example, temperature, pressure, or magnetic field. Early on, Lee and Yang realized that this non	analytic behavior can be understood by considering the zeros of the partition function in the complex plane of the control parameter~\cite{Yang1952a,Lee1952,Blythe2002,Bena2005}. In the case that a system exhibits a phase transition, the complex partition function zeros will approach the critical value on the real axis in the thermodynamic limit and thereby give rise to a nonanalytic behavior of the free-energy density~\cite{Fisher1978,Kenna1994,Biskup2000,Arndt2000,Kim2004,Ghulghazaryan2007,Lee2013,Wei2014,Kenna2014,GarciaSaez2015,Krasnytska2015a,Gnatenko2017,Kuzmak2019,Giordano2020,Matsumoto2022}. Lee and Yang thereby provided a rigorous foundation of phase transitions. Moreover, in recent years, it has been realized that partition-function zeros are not only a theoretical concept. They can also be determined in experiments~\cite{Binek1998,Wei2012,Peng2015,Flindt2013,Brandner2017}. In the approach that we follow here, the partition-function zeros can be found from the fluctuations of the thermodynamic observable that couples to the control parameter; for example, energy is coupled to the inverse temperature, while the magnetization of a spin lattice couples to the magnetic field~\cite{Flindt2013,Brandner2017,Deger2018,Deger2019,Deger2020,Deger2020b}. Thus, from these fluctuations, one can find the partition-function zeros for a given system size, and by gradually increasing the system size, one may determine the thermodynamic convergence point by extrapolation. This methodology has already been realized experimentally~\cite{Brandner2017}, and it has also been applied in theory to a variety of equilibrium problems, including simple models of DNA unfolding~\cite{Deger2018} and spontaneous magnetization in spin lattices~\cite{Deger2019,Deger2020,Deger2020b}. Surprisingly, in many cases, the critical value of the control parameter can be found using rather small system sizes. The framework is not restricted to equilibrium settings and can also be applied to non-equilibrium phase transitions~\cite{Flindt2013,Brange2022}. In addition, it was recently extended to the quantum realm to describe quantum phase transitions in the many-body ground state of interacting quantum systems~\cite{Kist2021,Vecsei2022,Vecsei2023} and to dynamical quantum phase transitions in spin lattices following a quench~\cite{Peotta2021,Brange2022b}.

The purpose of the present work is to apply the Lee-Yang methodology to predict the condensation temperature of a Bose gas from the fluctuations of the energy in a harmonic potential in one, two, and three dimensions, as illustrated in Fig.~\ref{Fig1}(a). Below the condensation temperature, the fraction of particles in the single-particle ground state grows algebraically as the temperature is reduced, as shown in Fig.~\ref{Fig1}(b) for a two-dimensional trapping potential. The abrupt change in the ground state population occurs in the thermodynamic limit of many particles. By contrast, we here consider fewer than 100 bosons, and we determine the partition-function zeros in the complex plane of the inverse temperature from the high cumulants of the energy fluctuations. Examples are provided in Fig.~\ref{Fig1}(c), where we show the complex partition-function zeros obtained with a high-temperature expansion of the partition function in one, two, and three dimensions. In two and three dimensions, the zeros converge towards the inverse transition temperature on the real axis as the number of particles is increased. On the other hand, for the one-dimensional trapping potential, there is no phase transition, and the zeros remain complex. 

This paper is organized as follows. In Sec.~\ref{Sec2}, we provide a brief overview of the standard approach to Bose-Einstein condensation based on the grand canonical ensemble. We then turn to the canonical ensemble, which we will use throughout this work, and where the number of particles is fixed. In Sec.~\ref{Sec3}, we discuss the Lee-Yang theory that we will be using, which considers the zeros of the canonical partition function in the complex plane of the inverse temperature. We show how the partition-function zeros can be determined from the energy fluctuations of the Bose gas and how we can predict the transition temperature in the thermodynamic limit using fewer than 100 bosons. In Sec.~\ref{Sec4}, we show results for bosons in a harmonic trap in one, two, and three dimensions, for which we predict the transition temperature by extrapolation to the thermodynamic limit. In Sec.~\ref{Sec5}, we then simulate the energy fluctuations in a Bose gas to demonstrate how the partition-function zeros in principle could be determined experimentally. Finally, in Sec.~\ref{Sec6}, we present our conclusions and provide an outlook on possible developments for the future. Additional technical details are provided in Appendixes~\ref{app;finiteBEC} and \ref{App:Derivation of Tc for 3d}. 

\section{Bose-Einstein condensation}
\label{Sec2}

\subsection{Grand-canonical ensemble}
The simplest approach to Bose-Einstein condensation is provided by the grand-canonical ensemble as discussed in most textbooks \cite{Annett2004}. The equilibrium properties of the gas are encoded in the grand-canonical partition function,
\begin{equation}
  \mathcal{Z}(\beta,\mu) = \mathrm{Tr}\left\{e^{-\beta(\hat{H}-\mu\hat{N})}\right\},
\end{equation}
where $\hat{H}$ is the many-body Hamiltonian of the confined bosons, the number operator is denoted by $\hat{N}$, and $\mu$ is the chemical potential, while $\beta=1/k_BT$ is the inverse temperature. The grand potential correspondingly reads
\begin{equation}
    \begin{split}
  \Phi(\beta,\mu) =& -\beta^{-1} \ln \mathcal{Z}(\beta,\mu)\\
  \simeq&-\beta^{-1} \int_0^\infty dEg_d(E)\ln\{1-e^{-\beta (E-\mu)}\},
  \end{split}
  \label{Grand potential}
\end{equation}
where 
\begin{equation}
g_d(E)=\frac{1}{(d-1)!}\frac{E^{(d-1)}}{(\hbar\omega_0)^d}
\end{equation}
is the density of states of the potential with frequency $\omega_0$ in $d$ dimensions. The mean number of bosons reads
\begin{equation}
    \begin{split}
  N =-\partial_\mu \Phi(\beta,\mu) = \int_0^\infty dE\frac{g_d(E)}{1-z e^{\beta E}}=\frac{\mathrm{Li}_d(z)}{(\beta\hbar\omega_0)^d},
  \end{split}
  \label{Tc for 2d and 3d}
\end{equation}
where $z=e^{\beta\mu}$ is the fugacity and $\mathrm{Li}_d(z)$ is the polylogarithmic function. At high temperatures, the contribution for the single-particle ground state can be included in the integrals above. On the other hand, at lower temperatures, the ground-state population becomes macroscopic and must be treated separately. When this happens, the chemical potential approaches the ground-state energy, which is set to zero, and we can thus find the condensation temperature from Eq.~\eqref{Tc for 2d and 3d} in the limit $z \to 1$, where the polylogarithmic function is given by Riemann's zeta function, $\mathrm{Li}_d(1)=\zeta(d)$, with $\zeta(2)=\pi^2/6$ and $\zeta(3)\simeq 1.2$, while it diverges for $d=1$. We then find the known expression for the transition temperature in $d=2,3$ dimensions,
\begin{equation}
T_c=\frac{\hbar\omega_0}{k_B}\left(\frac{N}{\zeta(d)}\right)^{1/d}.
\label{eq:tc_grand}
\end{equation}
By contrast, there is no phase transition for the one-dimensional harmonic potential. (See, however, Refs.~\cite{Ketterle1996,Mullin1997} and Appendix~\ref{app;finiteBEC} for a 
discussion of Bose-Einstein condensation of a finite number of trapped bosons in one dimension, which is not an actual phase transition in the thermodynamic limit.) The transition temperature in Eq.~(\ref{eq:tc_grand}) appears at first to diverge with the number of particles. However, in the thermodynamic limit, the particle density should be fixed, and to this end one weakens the potential, so that $\omega_0^d\times N$ is kept constant~\cite{Mullin1997}.

\subsection{Canonical ensemble}
In the approach that we pursue in the following, we consider the canonical ensemble, where the number of particles is fixed, and the partition function is defined as
\begin{equation}
  Z_N(\beta) = \mathrm{Tr}_N\left\{e^{-\beta\hat{H}_N}\right\}
\end{equation}
in terms of the Hamiltonian $\hat{H}_N$ for $N$ particles. Importantly, for an ideal Bose gas, the partition function can be found from the recursive expressionthe recursive expression~\cite{Schmidt1998,SCHMIDT1999,doi:10.1119/1.1544520,Borrmann2000,Mulken2001,Dijk2015}
\begin{equation}
    Z_N(\beta) = \frac{1}{N}\sum_{k=1}^N Z_1(k\beta) Z_{N-k}(\beta),
    \label{eq:recur}
\end{equation}
which starts with the partition function for a single particle together with $Z_0(\beta)=1$. For a single particle in a harmonic trapping potential, we readily find
\begin{equation}
  Z_1(\beta) = \left(\sum_{n=0}^\infty e^{-\beta \hbar \omega_0(n+1/2)}\right)^d = \left(\frac{e^{-\beta \hbar \omega_0/2}}{1-e^{-\beta\hbar\omega_0}} \right)^d
  \label{eq:Z1}
\end{equation}
and we can then obtain the partition function for $N$ particles using Eq.~(\ref{eq:recur}). At high temperatures, the partition function for a single particle simplifies to
\begin{equation}
Z_1(\beta)\simeq \frac{1}{(\beta\hbar\omega_0)^d},\quad \beta\hbar\omega_0\ll 1,
\label{eq:highT}
\end{equation}
which will be useful for some of our calculations.

\subsection{Moments and cumulants of the energy}

From the partition function and the associated free energy, we can obtain various thermodynamic observables and their moments and cumulants. In particular, we can obtain the moments of the total energy as
\begin{equation}
\label{eq:moments}
\langle U^n \rangle = (-1)^n \frac{\partial_\beta^nZ_N(\beta)}{Z_N(\beta)}=\mathrm{Tr}_N\left\{\hat{H}_N^n\frac{ e^{-\beta\hat{H}_N}}{Z_N(\beta)}\right\}.
\end{equation}
Similarly, the cumulants of the energy are defined as
\begin{equation}
\cum{U}{n} = (-1)^n \partial^n_{\beta} \ln{Z_N(\beta)},
\label{How to obtain the cumulants}
\end{equation}
and they can be directly expressed in terms of the moments via the standard relation
\begin{equation}
\cum{U}{n} = \langle U^n \rangle - \sum_{m=1}^{n-1} \binom{n-1}{m-1} \cum{U}{m} \langle U^{n-m} \rangle.
\end{equation}
Moreover, differentiating Eq.~\eqref{eq:recur} with respect to the inverse temperature and using the general Leibniz rule for the $n$th derivative of a product of functions, we arrive at
\begin{equation}\label{eq:derivative}
Z_N^{(n)}(\beta)=\frac{1}{N} \sum^N_{k=1} \sum^N_{l=0} \binom{n}{l} k^l Z_1^{(l)}(k\beta)Z_{N-k}^{(n-l)}(\beta),
\end{equation}
where we have introduced the notation $Z_N^{(n)}(\beta)=\partial_\beta^n Z_N(\beta)$. Using this expression, we can recursively evaluate the high derivatives of the partition function and the corresponding moments and cumulants using only the derivatives of the one-particle partition function.

\begin{figure*}[t]
    \centering
    \includegraphics[width=0.97\textwidth]{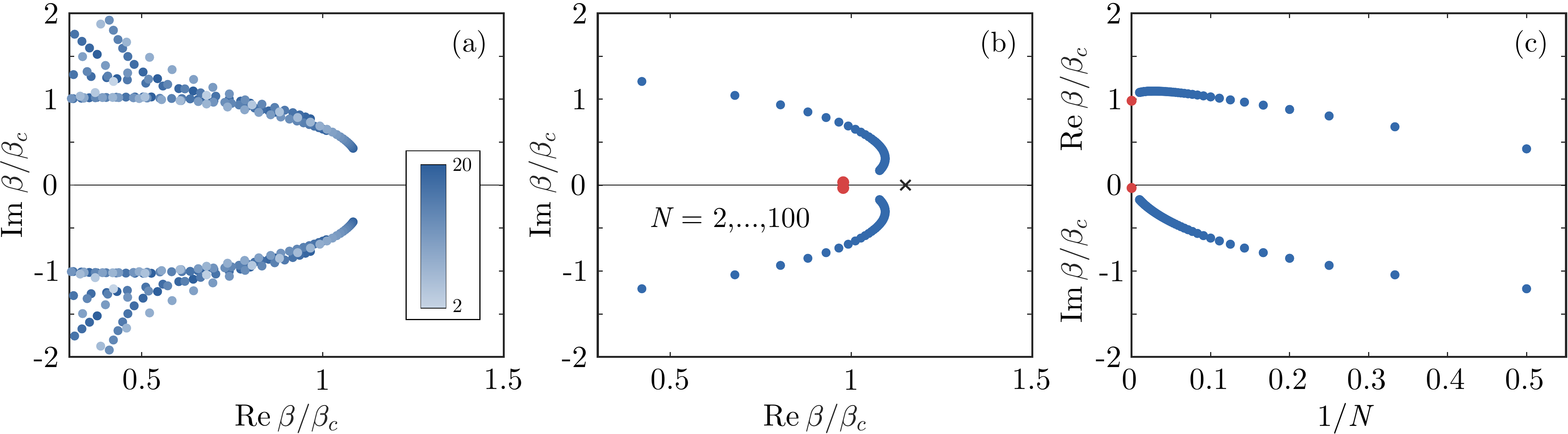}
    \captionsetup{justification=justified,singlelinecheck=false}
        \caption{Three-dimensional harmonic potential. (a) Partition-function zeros in the complex plane of the inverse temperature obtained numerically for $N=2,\ldots,20$ bosons. (b) Partition-function zeros closest to the inverse temperature $\beta = 1.15\beta_c$, marked with a cross, obtained with the cumulant method for $N=2,\ldots, 100$ bosons. Here, we have used $m=7$ different cumulant orders up to order $17$. (c) The real and (negative) imaginary parts of the zeros as a function of $1/N$. The red dots are the extrapolated convergence points in the thermodynamic limit, which we determine based on the scaling ansatz in Eq.~(\ref{eq:scaling}).}
    \label{Fig2}
\end{figure*}

\section{Lee-Yang theory}
\label{Sec3}
\subsection{Partition function zeros and poles}
In the works by Lee and Yang, they analyzed the nonanalytic behavior of the free-energy density at a phase transition in terms of the complex partition function zeros~\cite{Yang1952a,Lee1952}. In particular, they showed how a phase transition occurs as the zeros approach the real axis in the thermodynamic limit. In the case of Bose-Einstein condensation, the zeros will converge towards the value of the inverse temperature for which the phase transition occurs. In Ref.~\cite{Dijk2015}, the complex partition-function zeros of a Bose gas in a three-dimensional trapping potential were found numerically. Here, by contrast, we consider trapped bosons in one, two, and three dimensions, and we use a cumulant method to determine the partition-function zeros from the fluctuations of the energy, which in principle are measurable. We also discuss the fact that the partition function does not have only complex zeros. It also has poles, which is already clear from the single-particle partition function in Eq.~(\ref{eq:Z1}), which has poles along the imaginary axis, including one at the origin. These poles appear because of the infinitely many oscillator states in the sum in Eq.~(\ref{eq:Z1})~\cite{Brange2022,PhysRevB.99.085418}. 

\begin{figure*}[t]
    \centering
    \includegraphics[width=0.97\textwidth]{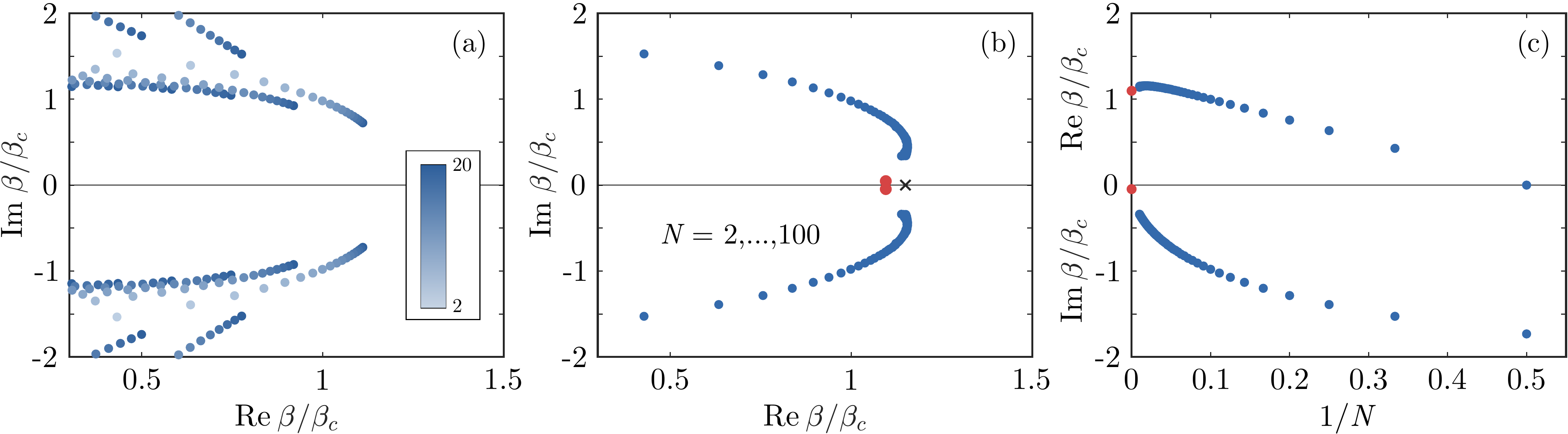}
    \captionsetup{justification=justified,singlelinecheck=false}
        \caption{Two-dimensional harmonic potential. (a) Partition function zeros in the complex plane of the inverse temperature obtained numerically for $N=2,\ldots,20$ bosons. (b) Partition function zeros closest to the inverse temperature $\beta = 1.15\beta_c$, marked with a cross, obtained with the cumulant method for $N=2,\ldots, 100$ bosons. Here, we have used $m=7$ different cumulant orders up to order $17$. (c) The real and (negative) imaginary parts of the zeros as functions of $1/N$. The red dots are the extrapolated convergence points in the thermodynamic limit, which we determine based on the scaling ansatz in Eq.~(\ref{eq:scaling}).}
    \label{Fig3}
\end{figure*}

As an illustration, we show in Fig.~\ref{Fig1}(c) the partition-function zeros in the complex plane of the inverse temperature, calculated numerically based on the recursive relation in Eq.~(\ref{eq:recur}) for $N=20$ particles together with the high-temperature expansion in Eq.~(\ref{eq:highT}). These results already indicate that Bose-Einstein condensation happens in the thermodynamic limit for two- and three-dimensional trapping potentials, while the zeros stay off of the positive real axis for a one-dimensional trap. More accurate results are shown in Figs.~\ref{Fig2}(a) and \ref{Fig3}(a), where we show the partition-function zeros for $N=2,\dots,20$ particles in two and three dimensions obtained with the exact expression for the single-particle partition function in Eq.~(\ref{eq:Z1}). These results are also indicative of a phase transition, although it is hard to reach larger system sizes as the calculations become increasingly cumbersome with increasing particle number. Instead, we make use of a cumulant method that allows us to determine the partition-function zeros from the fluctuations of the energy. Here, the starting point is a formal product expansion of the canonical partition function reading~\cite{Arfken2012a}
\begin{equation}
  Z_N\left(\beta\right) = c_0 e^{\beta c_1} \frac{\displaystyle \prod_{i} \left(\beta-\beta_i\right) }{\displaystyle \prod_{j}\left(\beta-\beta_j\right)},
  \label{Expansion of the partition function}
\end{equation}
where $c_0$ and $c_1$ are constants, and $\beta_i$ and $\beta_j$ are the zeros and poles of the partition function, respectively. For a finite number of particles, the zeros and the poles come in complex conjugate pairs, since the partition function is real valued for real values of $\beta$. 

\subsection{Cumulant method}

We now describe the cumulant method for extracting several of the partition-function zeros from the fluctuations of the energy. To this end, we use the definition of the cumulants in Eq.~(\ref{How to obtain the cumulants}) together with the product expansion in Eq.~\eqref{Expansion of the partition function} to express the cumulants as
\begin{equation}
\label{Expression for the cumulants in terms of singular points}
\langle\!\langle U^n\rangle\!\rangle = (n-1)!\sum_{k} \frac{(-1)^{p_k}}{(\beta-\beta_k)^n}, \quad n>1,
\end{equation}
where the sum runs over all zeros ($p_k=1$) and poles ($p_k=0$), which we collectively denote by $\beta_k$. Importantly, the contribution from each zero or pole decreases with the distance to the actual inverse temperature, $\beta$, to the power of the cumulant order. Thus, for high cumulant orders, we can truncate the sum as

\begin{equation}
\langle\!\langle U^n\rangle\!\rangle \simeq (n-1)! \sum_{k=1}^M \frac{(-1)^{p_k}}{(\beta-\beta_k)^n}, \quad n\gg 1,
\label{Approximated cumulants}
\end{equation}
where we have included only the $M$ zeros and poles that are closest to the actual inverse temperature. In the case with $M=2$, only the pair of zeros that are closest to the actual inverse temperature is included, and using the method developed in Refs.~\cite{Flindt2013,Brandner2017,Deger2018,Deger2019,Deger2020,Deger2020b}, one may extract these zeros from the high cumulants of the energy fluctuations. Recently, this approach was extended, so that more zeros and poles can be included in the sum and determined from the energy fluctuations~\cite{Peotta2021}. To simplify the notation, we define the normalized cumulants as
\begin{equation}
u_n = \frac{\langle\!\langle U^n\rangle\!\rangle}{(n-1)!},
\end{equation}
and then rewrite the truncated sum in Eq.~(\ref{Approximated cumulants}) as
\begin{equation}
u_n \simeq -\sum_{k=1}^m d_k \lambda_k^n
\label{eq:u_n}
\end{equation}
with
\begin{equation}
\lambda_k = \frac{1}{\beta-\beta_k}
\label{eq:lambdas}
\end{equation}
and $d_k$ being the multiplicity of each zero or pole (a simple pole has $d_k=-1$). To determine the zeros and poles in Eq.~(\ref{eq:lambdas}), we note that   expression \eqref{eq:u_n} for the normalized cumulants can be regarded as the general solution of a homogeneous difference equation of the form 
\begin{equation}
    u_{n} = a_1 u_{n-1} +a_2u_{n-2} +...+a_m u_{n-m}, 
    \label{Difference equation for cumulants}
\end{equation}
where the coefficients $a_k$ are still undetermined. To find them, we formulate a linear system of $m$ equations, using Eq.~(\ref{Difference equation for cumulants}), which reads
\begin{widetext}
\begin{equation}
\label{eq:aj_sys}
\begin{pmatrix}
    u_{n-1} & u_{n-2}  & \dots &  u_{n-m+1}  &  u_{n-m}  \\
    u_{n} & u_{n-1}  & \dots &  u_{n-m+2} &  u_{n-m+1}   \\
    \vdots & \vdots & \ddots & \vdots & \vdots \\
    u_{n+m-3} & u_{n+m-4}  & \dots & u_{n-1}  & u_{n-2} \\
     u_{n+m-2} & u_{n+m-3} & \dots & u_{n}   & u_{n-1}
\end{pmatrix}
\begin{pmatrix}
a_1 \\ a_2 \\ \vdots \\a_{m-1} \\a_{m}
\end{pmatrix} =
\begin{pmatrix}
u_{n}   \\ u_{n+1}   \\ \vdots \\ u_{n+m-2}   \\ u_{n+m-1} 
\end{pmatrix}.
\end{equation} 
Recalling that the normalized cumulants $u_k$ are known, we may now solve for the coefficients $a_k$ by inverting the matrix on the left-hand side of this equation. Afterwards, we may find $\lambda_k$ as the roots of the equation
\begin{equation}
    \lambda^m-a_1\lambda^{m-1}-a_2\lambda^{m-2}-...-a_{m-1}\lambda - a_m = 0,
\end{equation}
which is the characteristic equation obtained from the recurrence relation in Eq.~(\ref{Difference equation for cumulants}). As the last step, we obtain the multiplicities $d_k$ as the solutions to the matrix equation
\begin{equation}
-\begin{pmatrix}
    1         & 1         & \dots & 1   & 1 \\
    \lambda_0 & \lambda_1 & \dots & \lambda_{m-2} & \lambda_{m-1} \\
    \vdots & \vdots & \ddots & \vdots & \vdots \\
    \lambda_0^{m-2} & \lambda_1^{m-2} & \dots & \lambda_{m-2}^{m-2} & \lambda_{m-1}^{m-2} \\[0.3em]
    \lambda_0^{m-1} & \lambda_1^{m-1} & \dots & \lambda_{m-2}^{m-1} & \lambda_{m-1}^{m-1}
\end{pmatrix}
\begin{pmatrix}
d_0\lambda_0^n \\ d_1\lambda_1^n \\ \vdots \\d_{m-2}\lambda_{m-2}^n \\[0.3em]
 d_{m-1}\lambda_{m-1}^n
\end{pmatrix} =
\begin{pmatrix}
u_{n} \\ u_{n+1}  \\ \vdots \\ u_{n+m-2}  \\[0.3em]
u_{n+m-1} 
\end{pmatrix}\,,
\label{Equation for d coefficients}
\end{equation}
\end{widetext}
which follows directly from Eq.~(\ref{eq:u_n}).

From the roots, we immediately find the zeros and poles as $\beta_k = \beta-1/\lambda_k$ together with their multiplicities $d_k$ from the last matrix equation. Importantly, the accuracy of the method can be gauged by the multiplicities, which should be integers. Thus, deviations from integer values indicate that the truncation in Eq.~\eqref{Approximated cumulants} is not a good approximation. One can then include more cumulants or increase the cumulant order to ensure convergence. For the following results, we have carefully checked that the position of the zeros and the poles have converged with respect to the number of cumulants and their order. Also, we consider only zeros and poles whose multiplicity coefficients $d_k$ deviate from an integer value by less than a tolerance threshold of 0.3. 

\begin{figure*}[t]
    \centering
    \includegraphics[width=0.97\textwidth]{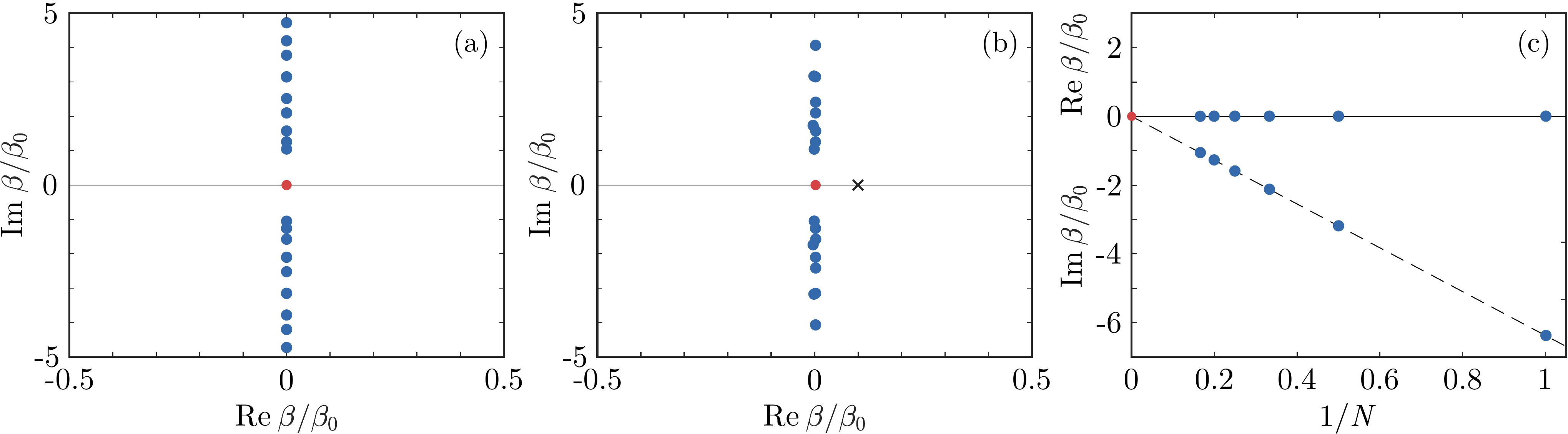}
    \captionsetup{justification=justified,singlelinecheck=false}
        \caption{One-dimensional harmonic potential. (a) Poles of the partition function according to Eq.~(\ref{eq:poles}) for $N=6$ bosons with $\beta_0=1/(\hbar\omega_0)$. (b) Poles closest to the origin obtained with the cumulant method for $N=2,\ldots, 6$ bosons. Here, we have used $m=6$ different cumulant orders up to order $25$, with a tolerance threshold of 0.25. (c) The real and (negative) imaginary parts of the poles a function of $1/N$ together with a linear extrapolation shown by a dashed line.}
    \label{Fig4}
\end{figure*}
 
\section{Condensation temperature}
\label{Sec4}
\subsection{Three-dimensional trapping potential}
In Fig.~\ref{Fig2}, we show partition-function zeros in the complex plane of the inverse temperature for the three-dimensional harmonic trapping potential. In Fig.~\ref{Fig2}(a), we show numerical results for the partition-function zeros using system sizes in the range $N=2,\ldots,20$. By contrast, in Fig.~\ref{Fig2}(b), we show results obtained using the cumulant method, which allows us to extract the conjugate pair of zeros that are closest to the inverse temperature at which the cumulants are evaluated, marked with a cross on the real axis. Thus, from the energy fluctuations of the Bose gas at this inverse temperature, we can determine the closest partition-function zeros and monitor their motion as the system size is gradually increased in the range $N=2,\ldots,100$. From the analysis of Bose-Einstein condensation in the grand-canonical ensemble, we know that the critical inverse temperature depends on the number of particles according to Eq.~(\ref{eq:tc_grand}). For this reason, we use dimensionless units on the axes, where $\beta_c=1/(k_BT_c) $ is the expected inverse transition temperature. Here, we note that there are finite-size corrections to the predicted transition temperature, which can be obtained from the equation (see Appendix~\ref{App:Derivation of Tc for 3d} for details)
\begin{equation}
T_{c}=\frac{\hbar\omega_0}{k_B}\left(\frac{N}{\zeta(3)}\right)^{1/3}\left[1+\frac{3}{2}\frac{\zeta(2)}{\zeta(3)}\frac{\hbar\omega_0}{k_BT_c}\right]^{-1/3}. \label{eq:TcFinite}
\end{equation}
However, in the thermodynamic limit, where $\omega_0^3N$ is kept constant by weakening the potential, we see that this expression reduces to the one in Eq.~(\ref{eq:tc_grand}) for $d=3$.

To determine the convergence point in the thermodynamic limit, we show in Fig.~\ref{Fig2}(c) the real and (negative) imaginary parts of the zeros as a function of the inverse particle number. We then use the scaling ansatz~\cite{Deger2019,Deger2020,Deger2020b} \begin{equation}
    |\beta_0-\tilde\beta_c|\propto N^{-\alpha}
    \label{eq:scaling}
\end{equation}
from which we find $\alpha$ and the prediction for the critical point $\tilde\beta_c$, marked with red circles, with the method of least squares. These results show how the extrapolated convergence points come close to the expected inverse transition temperature with only a small imaginary part. Thus, we see how the transition temperature in the thermodynamic limit can be rather precisely estimated from the energy fluctuations of a gas with fewer than 100 bosons.

\subsection{Two-dimensional trapping potential}
In Fig.~\ref{Fig3}, we show results for the two-dimensional harmonic trapping potential. In Fig.~\ref{Fig3}(a), we again show numerical results for the partition-function zeros using system sizes in the range $N=2,\ldots,20$. In addition, we show in Fig.~\ref{Fig3}(b) results obtained using the cumulant method based on cumulants of the energy evaluated at the inverse temperature marked with a cross on the real axis. To determine the convergence points of the partition function zeros, we again use the scaling ansatz in Eq.~(\ref{eq:scaling}) and find the extrapolated values indicated with red circles in Fig.~\ref{Fig3}(b) and \ref{Fig3}(c). Similar to the three-dimensional trapping potential, the extrapolated convergence point is in good agreement with the expected inverse transition temperature. However, we notice that the convergence of the zeros to the real axis is slower than for the three-dimensional case. Thus, the estimation of the critical temperature is not as accurate for the two-dimensional case as for the three-dimensional case.

\subsection{One-dimensional trapping potential}
The results for the harmonic potential in two and three
dimensions illustrate how the transition temperature can be determined from the energy fluctuations of fewer than 100 particles. As our last application, we consider bosons in a one-dimensional harmonic trapping potential. In this case, no phase transition is expected in the thermodynamic limit. In particular, the partition function takes on the simple expression \cite{doi:10.1119/1.1544520}
\begin{equation}
    Z_N(\beta) = e^{\beta\hbar\omega_0 N(N-1)/4} \prod_{k=1}^N Z_1(\beta k),
\end{equation}
where $Z_1(\beta)$ is the partition function for a single particle in Eq.~(\ref{eq:Z1}). We then see that the partition function for $N$ particles only has poles along the imaginary axis at
\begin{equation}
    \beta_{k,n} = \frac{2\pi}{\hbar\omega_0}\frac{n}{k}i, 
    \,\,\,\begin{array}{l} 
    n=\ldots,-1,0,1,\ldots \\
    k=1,\ldots,N
    \end{array}
    \label{eq:poles}
\end{equation}
as shown in Fig.~\ref{Fig4}(a).

\begin{figure*}
    \centering
    \includegraphics[width=0.97\textwidth]{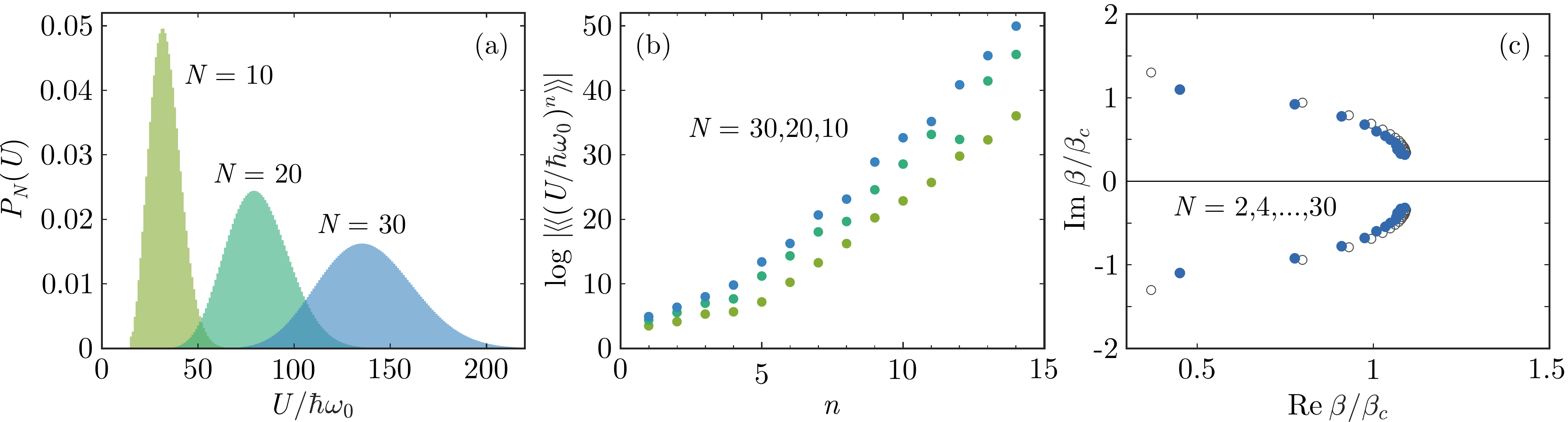}
    \captionsetup{justification=justified,singlelinecheck=false}
        \caption{Partition-function zeros from energy fluctuations. (a) Distributions $P_N(U)$ of the energy for a three-dimensional harmonic potential with $N=10, 20, 30$ particles. (b) Energy cumulants of orders $n=1,...,14$, obtained from $10^7$ Monte Carlo simulations. (c) Partition-function zeros obtained by averaging over 1000 sets of Monte Carlo simulations are shown by solid circles. The exact zeros are shown with open circles. For all of these calculations, the inverse temperature is $\beta = 1.15\beta_c$.}
    \label{Fig5}
\end{figure*}

In Fig.~\ref{Fig4}(b), we show the poles that are closest to the origin obtained with the cumulant method for $N=2,\ldots, 6$ bosons. We focus on the poles that are not exactly at the origin and have thus multiplied the partition function in Eq.~\eqref{eq:Z1} by $\beta$ to eliminate the pole at $\beta=0$. As expected, we find no zeros or poles away from the imaginary axis in the complex plane of the inverse temperature, which implies that there is no phase transition in the thermodynamic limit. In Fig.~\ref{Fig4}(c), we investigate the real and (negative) imaginary parts of the poles as a function of the inverse particle number to extrapolate their position in the thermodynamic limit. In agreement with Eq.~(\ref{eq:poles}), we find that the pair of poles that are closest to the origin converge towards the origin as the number of bosons is increased. With this example we see that the cumulant method correctly predicts that there is no phase transition in the thermodynamic limit.

\section{Energy fluctuations}
\label{Sec5}
Finally, we show how the zeros can be determined from simulated fluctuations of the energy, which we generate using a Monte Carlo method (and which, in principle, are measurable). For the simulations, we need the probability $P_N(U_m)$ that $N$ bosons in $d$ dimensions have the energy $U_m = (m+Nd/2)\hbar \omega_0$ with $m=0,1,2,\dots$. To this end, we write the  partition function as
\begin{equation}
    Z_N(\beta)=\sum_{m=0}^\infty G_m e^{-\beta U_m},
\end{equation}
where the degeneracy $G_m$ enters the probabilities as
\begin{equation}
    P_N(U_m)=\frac{G_m e^{-\beta U_ m}}{Z_N(\beta)}.
    \label{eq:prob}
\end{equation}
We also renormalize the partition function as
\begin{equation}
z_N(\beta)=e^{\beta\hbar\omega_0 Nd/2}Z_N(\beta),
\label{eq:znrenorm}
\end{equation}
and can then define the probability-generating function
\begin{equation}
    \mathcal{G}_N(v)= \frac{z_N(\beta-\ln v/\hbar\omega_0)}{z_N(\beta)}=\sum_m P_N(U_m)v^m,
\end{equation}
having introduced the variable $v=e^{-\beta\hbar\omega_0}$. The generating function can be inverted for the probabilities as
\begin{equation}
 P_N(U_m) = \frac{1}{2\pi i}\oint_{\mathcal C} dv\frac{\mathcal{G}_N(v)}{v^{m+1}},
\label{eq:cauchy}
\end{equation}
where $\mathcal C$ is a positively oriented circle that is centered around the origin with a radius that is smaller than 1. To find the probabilities in a numerically stable manner, we use the scheme from Refs.~\cite{ABATE1992245,Kambly2014} for the inversion of probability generating functions. We then have
\begin{equation}
\label{Estimated probability distribution}
\begin{split}
P_N(U_m) & \simeq \frac{1}{2mr^m} \Big(r\left[\mathcal{G}_N(r)-(-1)^m  \mathcal{G}_N(-r)\right] \\
&+2\sum_{j=1}^{m-1}(-1)^j \textrm{Re}[re^{ij\pi/m}\mathcal{G}_N(re^{ij\pi/m})] \Big), 
\end{split}
\end{equation}
where $0<r<1$ is a small parameter, which ensures that deviations from the exact values are smaller than $r^{2m}/(1-r^{2m})$~\cite{ABATE1992245}. We use $r = 10^{-10/(2m)}$ in all calculations, so that the deviations are smaller than $10^{-10}$.

In Fig.~\ref{Fig5}(a), we show energy distributions for different numbers of particles in a three-dimensional trapping potential.  As one would expect, the distributions shift towards higher energies as the number of particles is increased. We use these distributions as the starting point for our Monte Carlo simulations, where we simulate $10^7$ measurements of the energy. From these simulations, we then obtain the high cumulants in Fig.~\ref{Fig5}(b). Next, using the cumulant method, we extract the partition function zeros, shown with solid circles in Fig.~\ref{Fig5}(c). These zeros were obtained by averaging over 1000 sets of simulations. For the sake of comparison, we also show the exact zeros with open circles, and we see that the Monte Carlo simulations agree well with the exact results. We note that cumulants of very high orders have been measured for electron tunneling through quantum dots~\cite{Brandner2017,Flindt2009}. 

\section{Conclusions and outlook}
\label{Sec6}
In conclusion, we have presented a Lee-Yang theory of Bose-Einstein condensation and used a cumulant method to predict the condensation temperature from the energy fluctuations of fewer than 100 bosons. For harmonic trapping potentials in two and three dimensions, we obtained predictions of the condensation temperature in the thermodynamic limit of many particles and large volumes, which agree well with known results. Moreover, for a one-dimensional trapping potential, we found no phase transition in the thermodynamic limit as one would expect. Our approach is directly related to observables that, in principle, are measurable, namely the fluctuations of the energy, and it thereby provides a link between experimental observations and the Lee-Yang theory of phase transitions. We have illustrated our methodology with noninteracting particles. However, our approach would work equally well for interacting particles as long the fluctuations of the energy are accessible. Potentially, with further refinements, our method could also be applied to nonequilibrium Bose-Einstein condensates  \cite{Vorberg2013,Schnell2017,Schnell2018}.  The cumulant method has already been realized in experiments on electron tunneling~\cite{Brandner2017}, and one may hope that it can also be implemented for cold atoms with a small and controllable number of bosons.

\acknowledgements
This work was supported by the Academy of Finland through the Finnish Centre of Excellence in Quantum Technology (Projects No.~312057 and No.~312299) and
Grants No.~318937 and No.~331737. The work was also supported by the Medical Research Council (Grant No.~ MR/S034714/1) and the Engineering and Physical Sciences Research Council (Grant No.~EP/V031201/1). T.P.~acknowledges support from the Nokia Industrial Doctoral School in Quantum Technology. K.B.~acknowledges support from the University of Nottingham through a Nottingham Research Fellowship. 

\begin{figure*}[t]
    \centering
    \includegraphics[width=0.97\textwidth]{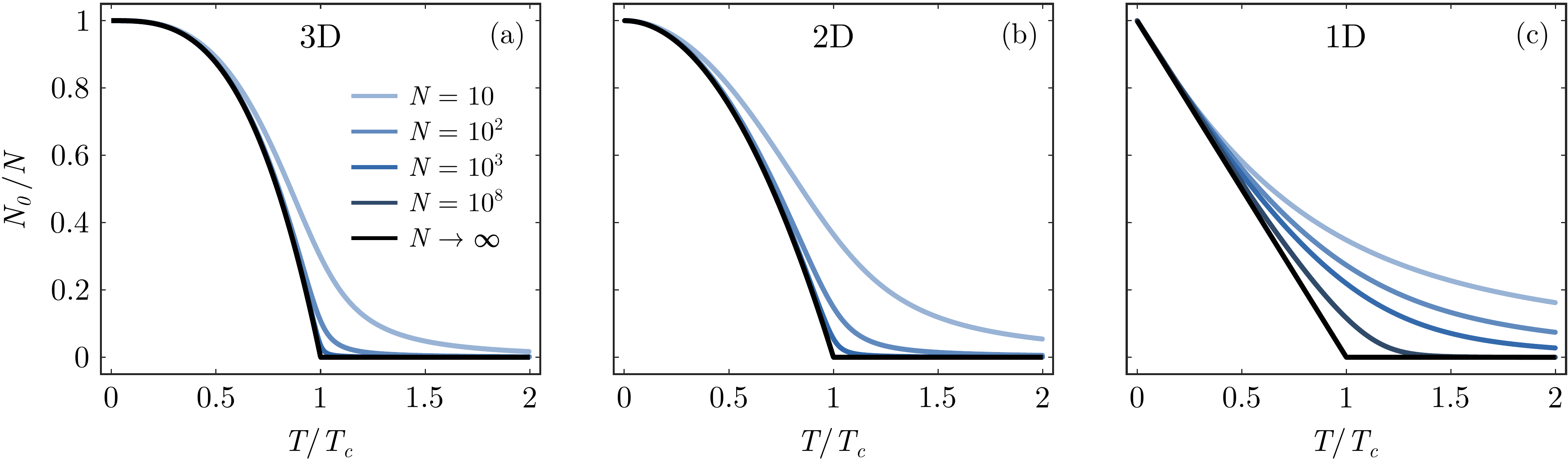}
    \captionsetup{justification=justified,singlelinecheck=false}
        \caption{Bose-Einstein condensation of a finite number of particles. We show the fraction of particles in the ground state as a function of the temperature and with different numbers of particles in the trapping potential. The three panels correspond to different dimensions, and the temperature $T_c$ depends on the number of particles according to Eqs.~(\ref{eq:tc_grand}), (\ref{eq:TcFinite}), and (\ref{eq:Tc1}). The black lines are given by Eq.~(\ref{eq:thermo_frac}), which holds in the thermodynamic limit. For all three dimensions, the fraction of particles in the ground state can become large with a finite number of particles. However, for the one-dimensional trapping potential, the transition temperature $T_c$ goes to zero in the thermodynamic limit, and there is no phase transition. }
    \label{Fig6}
\end{figure*}

\appendix
\section{Bose-Einstein condensation of a finite number of particles}
\label{app;finiteBEC}

Typically, phase transitions are associated with a nonanalytic behavior of the free energy, which emerges in the thermodynamic limit of many particles and large volumes. While Bose-Einstein condensation in two- and three-dimensional harmonic potentials is consistent with this picture, the case of a one-dimensional harmonic trapping potential is different, and no singularities appear in the thermodynamic limit. Nevertheless, as discussed in Refs.~\cite{Ketterle1996,Mullin1997}, for a finite number of particles the ground state may still become macroscopically occupied below a certain temperature, which resembles the behavior of the two- and three-dimensional potentials. To see this, we consider the grand-canonical ensemble and relate the average number of particles to the fugacity as
\begin{equation}
N = \frac{z}{1-z} + \frac{\mathrm{Li}_d(z)}{(\beta\hbar\omega_0)^d},
\label{eq:eqforz}
\end{equation}
where the first term is the occupation of the ground state. The fraction of particles in the ground state then becomes
\begin{equation}
\frac{N_0}{N} = \frac{1}{N}\frac{z}{1-z},
\label{fraction of particles}
\end{equation}
where the fugacity is obtained by solving Eq.~(\ref{eq:eqforz}) for the fugacity in terms of the particle number $N$. Moreover, in the thermodynamic limit, the fraction can be written as
\begin{equation}
\frac{N_0}{N} = 1-(T/T_c)^d,\,\,\, T\leq T_c,
\label{eq:thermo_frac}
\end{equation}
where $T_c$ for $d=2,3$ is given by Eq.~(\ref{eq:tc_grand}) and for $d=1$,
\begin{equation}
T_c=\frac{\hbar\omega_0}{k_B} \frac{N}{\ln N}.
\label{eq:Tc1}
\end{equation}
Equation~\eqref{eq:Tc1} follows from the assumption that the ground-state population $N_0$ is macroscopic, i.e., on the order of $N$, which leads to the chemical potential approaching the ground-state energy. Hence, we can expand Eqs.~(\ref{eq:eqforz}) and (\ref{fraction of particles}) around small values of $\beta\mu\ll 1$ and take the limit of large $N$ to arrive at Eqs.~(\ref{eq:thermo_frac})~and~(\ref{eq:Tc1}).

Importantly, the transition temperature remains finite in two and three dimensions since the product $\omega_0^d\times N$ is kept constant in the thermodynamic limit. By contrast, the transition temperature goes to zero for the one-dimensional potential because of the logarithmic term in the denominator. For this reason, there is no Bose-Einstein condensation for the one-dimensional harmonic potential in the thermodynamic limit. On the other hand, for a finite number of particles, the temperature in Eq.~(\ref{eq:Tc1}) is finite, and the fraction of particles in the ground state can become large as illustrated in Fig.~\ref{Fig6}. We note that similar figures can be found in Refs.~\cite{Ketterle1996,Mullin1997}.

\section{Finite-size correction to the transition temperature for the three-dimensional trap}
\label{App:Derivation of Tc for 3d}

The condensation temperature in Eq.~\eqref{eq:tc_grand} can be modified to account for the system having a finite number of particles. To this end, we note that the energy levels,
\begin{equation}
E =\hbar\omega_0(n_x+n_y+n_z+3/2), n_i=0,1,2\ldots,
\end{equation}
of the three-dimensional harmonic potential are degenerate with the degeneracy factors,
\begin{equation}
g_n = \frac{1}{2}(n+1)(n+2),
\end{equation}
where $n=n_x+n_y+n_z$. The average particle number can then be expressed as 
\begin{equation}
    N = \sum_{n=0}^\infty g_n N_n \simeq N_0 + \int_0^\infty dn \, g_n N_n, \label{eq:TotalParticleNumber}
\end{equation}
where $N_n = (e^{\beta\hbar\omega_0n}-1)^{-1}$ is a Bose-Einstein factor, and  $N_0 = \mathrm{Li}_0(z)$ is the ground-state population. Evaluating the integral, we find
\begin{equation}
    N = \mathrm{Li}_0(z) + \frac{\mathrm{Li}_1(z)}{(\beta\hbar\omega_0)} + \frac{3}{2} \frac{\mathrm{Li}_2(z)}{(\beta\hbar\omega_0)^2} + \frac{\mathrm{Li}_3(z)}{(\beta\hbar\omega_0)^3}, \label{eq:TotalParticles}
\end{equation}
where the last two terms, which involve only contributions from the excited states in the integral of Eq.~\eqref{eq:TotalParticleNumber}, dominate above the critical temperature. 
Ignoring the first two terms in Eq.~\eqref{eq:TotalParticles} and taking the limit $z \to 1$ as was done in Sec.~\ref{Sec2} to find the condensation temperature in the thermodynamic limit, we find an equation for $T_c$, which can be rearranged to arrive at Eq.~\eqref{eq:TcFinite}. We note that there is not a similar correction to the transition temperature for the two-dimensional trapping  potential.

%

\end{document}